\documentclass[journal]{IEEEtran}
\usepackage{amsmath,amsfonts}
\usepackage{algorithm2e}
\usepackage{array}
\usepackage[caption=false,font=normalsize,labelfont=sf,textfont=sf]{subfig}
\usepackage{textcomp}
\usepackage{stfloats}
\usepackage{url}
\usepackage{verbatim}
\usepackage{graphicx}
\usepackage{cite}

\usepackage[top=0.6in, bottom=0.72in, left=0.56in, right=0.56in]{geometry}
\usepackage{makecell}

\usepackage{graphicx} 
\usepackage{amsmath}
\usepackage{amsmath}
\usepackage{amsmath, amssymb}
\usepackage{xcolor}	

\usepackage{booktabs}

\usepackage{algpseudocode}
\usepackage{graphics}
\usepackage{amsmath, amsthm}

\usepackage[caption=false,font=normalsize,labelfont=sf,textfont=sf]{subfig}

\newcounter{assumption}

\newcounter{theorm} 

\begin{document}
	


    \title{Impact of Position Uncertainty on the Secrecy Performance of Pinching Antenna Systems
    

    }

	\author{IEEE Publication Technology,~\IEEEmembership{Staff,~IEEE,}
	}

	\author{Saeid Pakravan, Imene Trigui, Wessam Ajib, Wei-Ping Zhu 

\thanks{S. Pakravan and W. Ajib are with the Department of Computer Sciences, University of Quebec in Montreal (UQAM), Montreal, QC, Canada. email: pakravan.saeid@uqam.ca; ajib.wessam@uqam.ca.}

\thanks{I. Trigui is with the Department of Applied Sciences, University of Quebec at Chicoutimi (UQAC), Chicoutimi, QC, Canada. email: itrigui@uqac.ca.}

\thanks{W.-P. Zhu is with the Department of Electrical and Computer Engineering, Concordia University, Montreal, QC, Canada. email: weiping@ece.concordia.ca.}

}
	\maketitle

\begin{abstract}

This paper investigates the secrecy performance of pinching-antenna systems (PAS) under practical pinching-position activation uncertainty. By dynamically selecting the radiation point along a dielectric waveguide, PAS enables low-cost spatial reconfigurability and enhanced secure transmission. Unlike existing studies that assume ideal activation control, we account for spatial inaccuracies caused by hardware limitations and environmental perturbations, which induce statistical dependence between the legitimate and eavesdropping channels. To characterize this uncertainty-induced dependence, PAS-specific marginal signal-to-noise ratio (SNR) distributions are derived, and a Gaussian copula is employed as a tractable representation of the resulting joint SNR distribution, enabling the derivation of approximate expressions for the secrecy outage probability. Simulation results validate the theoretical findings and demonstrate that PAS retains robust secrecy performance compared with conventional fixed-antenna systems, even in the presence of activation uncertainty.


\end{abstract}


\section{Introduction}

The rapid expansion of intelligent mobile devices, mission-critical services, and location-aware applications has heightened the demand for secure, reliable, and flexible communication in beyond-fifth-generation (B5G) and 6G wireless networks \cite{8869705}. These networks are expected to support dense connectivity, low-latency, and robust security performance, particularly in challenging propagation conditions and cost-constrained deployments. While technologies such as massive multiple-input multiple-output, reconfigurable intelligent surfaces, and adaptive antenna architectures enhance spatial degrees of freedom and network controllability \cite{10506508, 10713386, 10906511, 11106811, 10278220}, their deployment is often limited by hardware complexity, power consumption, and physical design constraints. These limitations have motivated the exploration of alternative antenna architectures that enable spatial reconfigurability with reduced radio-frequency (RF) complexity.

In this context, pinching-antenna systems (PAS) have recently emerged as an attractive architectural paradigm \cite{10945421, suzuki2022pinching,10976621, 11481959}. A PAS employs a dielectric waveguide along which a small dielectric particle can be ``pinched,'' thereby creating a radiating point at any desired position. By activating this pinch at different locations, the system can flexibly reconfigure its effective antenna position without mechanically moving the entire antenna structure or deploying additional RF chains. Unlike other spatially reconfigurable architectures, such as movable antennas and fluid antenna systems \cite{10906511, 11106811, 10278220}, which rely on antenna displacement or aperture reconfiguration, PAS realizes spatial adaptability through waveguide-assisted pinching activation. This geometry-controlled operation enables flexible radiating-point placement with low RF hardware complexity and power consumption. Moreover, by reducing the propagation distance to the intended receiver, PAS can alleviate blockage effects, enhance LoS probability, and improve spatial diversity \cite{11016750}. These features make PAS attractive for lightweight and cost-efficient next-generation wireless platforms.

Motivated by these advantages, recent studies have explored PAS in a range of wireless communication scenarios. Initial works focused on performance gains achievable through dynamic antenna repositioning; for instance, \cite{10896748} showed that optimizing the activation point in a downlink PAS can significantly enhance user rate. Multi-user extensions in \cite{pinch1111} employed joint antenna positioning and resource allocation to improve system energy efficiency. PAS has also been investigated in emerging areas such as integrated sensing and communication, where its LoS-enhancing capability supports simultaneous sensing and data transmission \cite{pinchISAC}. In non-orthogonal multiple access systems, their reconfigurability enables adaptive spatial separation and improved interference management \cite{10912473, 11029492}. Moreover, recent studies on physical-layer security (PLS) \cite{pinchplc1, pinchplc2, 11215679} show that positioning the radiating point near the legitimate receiver can provide notable secrecy gains over conventional fixed-antenna designs.

Despite the recent progress in PAS-enabled wireless communications, existing studies on PAS-assisted PLS mainly rely on idealized assumptions, including perfect pinching-position activation and statistically independent legitimate and eavesdropping links. Under practical operation, however, hardware imperfections, localization inaccuracies, and environmental perturbations introduce uncertainty in the activated pinching position, which inherently couples the propagation conditions of Bob and Eve. Consequently, the independence assumption commonly adopted in existing PAS secrecy studies may no longer accurately characterize the underlying channel behavior.
To this end, we investigate the secrecy performance of PAS under pinching-position uncertainty. We show that activation-position uncertainty alters the statistical structure of the secrecy problem by inducing statistical dependence between the legitimate and eavesdropping links. To characterize this uncertainty-induced dependence, PAS-specific marginal signal-to-noise ratio (SNR) distributions are derived, while a Gaussian copula is employed within a dependence-aware secrecy framework to represent the resulting joint SNR distribution. Based on this framework, approximate expressions for the secrecy outage probability (SOP) are derived and validated through Monte Carlo (MC) simulations, thereby quantifying the impact of activation uncertainty on secrecy performance.


\section{System Model}

We consider the downlink of a secure wireless system consisting of a base station (BS), a legitimate user (Bob), and a passive eavesdropper (Eve), as shown in Fig.~1. Bob and Eve are equipped with single fixed-position antennas, while the BS employs a dielectric waveguide with a reconfigurable pinching radiator \footnote{The extension of the proposed framework to multi-user and multi-pinching-antenna architectures remains an interesting direction for future research.}. By dynamically activating a radiating point along the waveguide, the BS aims to enhance Bob's link quality relative to Eve's, thereby improving overall secrecy performance.

\begin{figure}[]
    \centering
\includegraphics[width=0.76\linewidth, height=0.35\linewidth]{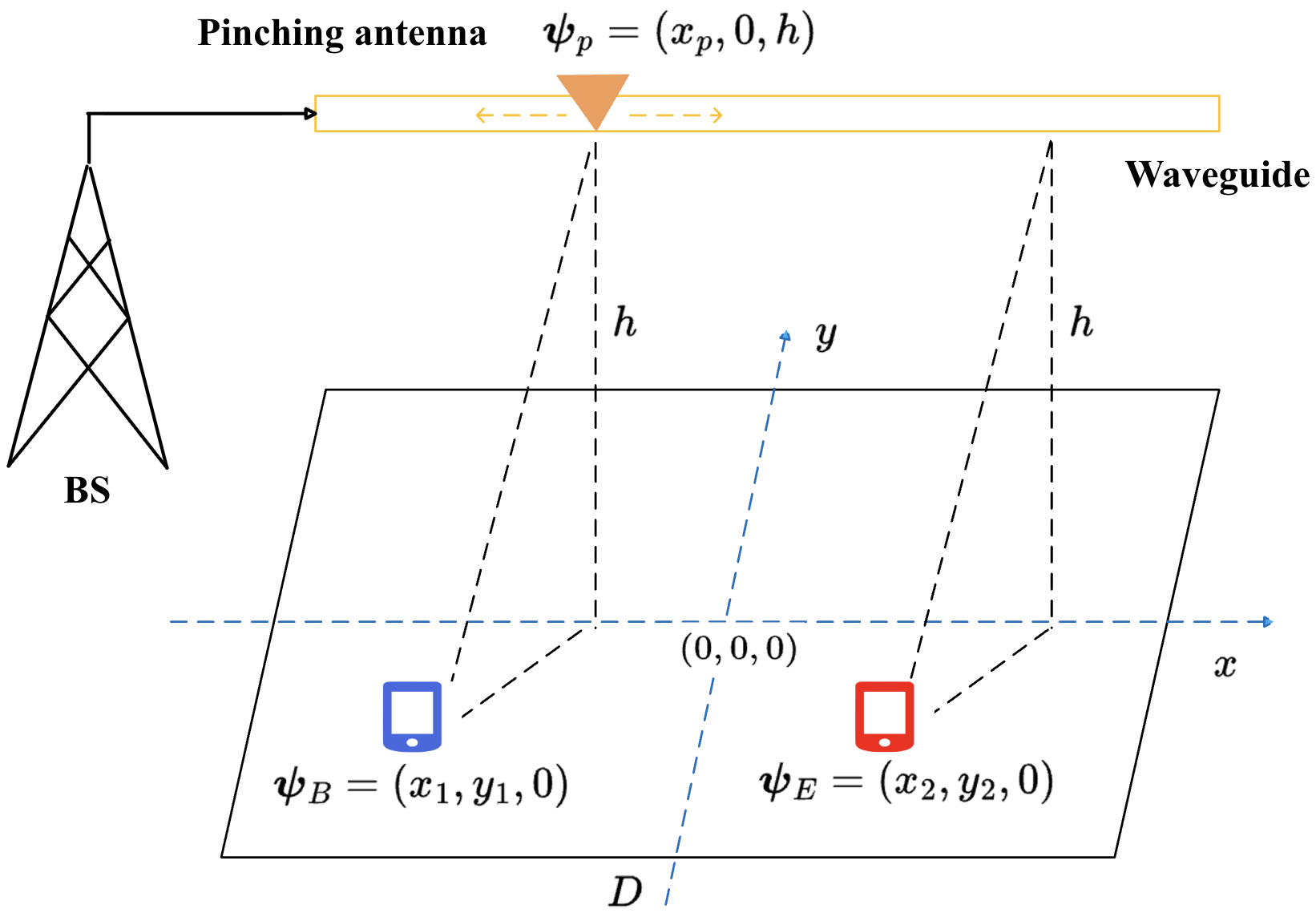}
    \caption{Secure PAS with pinching position uncertainty.}
    \label{fig:system_model}
\end{figure}

The waveguide is mounted parallel to the $x$--axis at height $h$, and the active pinching point has coordinates $\boldsymbol{\psi}_p = (x_p,0,h)$, where $x_p$ denotes the horizontal activation position \footnote{
In practice, feasible pinching locations may be discrete due to implementation constraints. The continuous-position model adopted herein serves as an analytical benchmark, whereas practical discretization may limit spatial reconfigurability and consequently affect secrecy performance.
}. Bob and Eve are randomly distributed within a square service area of side length $D$, with locations $\boldsymbol{\psi}_{{B}}=(x_1,y_1,0)$ and $\boldsymbol{\psi}_{{E}}=(x_2,y_2,0)$, respectively, where  \(x_i,y_i\sim\mathrm{Unif}[-D/2,D/2]\), \(i\in\{1,2\}\).  
The waveguide feed point  \(\boldsymbol{\psi}_0=(-D/2,0,h)\) has negligible impact on the large-scale secrecy analysis once the effective transmit SNR is defined. A deterministic LoS channel with free-space path loss is considered between the active pinching point and each receiver. This assumption is motivated by the high-frequency operating regime envisioned for PAS, where propagation is highly directional and the dominant received power is typically carried by the LoS component \cite{10945421, suzuki2022pinching,10976621, 6363891, 11481959}. Waveguide attenuation is absorbed into the effective transmit SNR, allowing the analysis to focus on the geometry-dependent secrecy behavior induced by pinching-position uncertainty. Since Eve is passive, the BS is assumed to have channel state information (CSI) only for Bob. Ideally, the pinching (radiating) point is activated horizontally above Bob, i.e., $x_p=x_1$, which maximizes his received signal strength, but in
practice, localization errors and hardware limitations introduce
positioning uncertainty, modeled as
\begin{equation}
    x_p = x_1 + E,
\end{equation}
where $E\sim \mathrm{Unif}[-\Delta,\Delta]$, $0 < \Delta \le D/2$, denotes the positioning error, independent of user locations. Thus, the active pinching point has coordinates $\boldsymbol{\psi}_p = (x_1 + E,0,h)$.

Using a spherical-wave propagation model \cite{10945421,10976621}, the baseband channel coefficient between the pinching point and a generic receiver located at $\boldsymbol{\psi}_u$ is expressed as
\begin{equation}
    h_u=
    \frac{\eta^{1/2}\exp\!\left(-j\frac{2\pi}{\lambda}\|\boldsymbol{\psi}_p-\boldsymbol{\psi}_u\|\right)}
    {\|\boldsymbol{\psi}_p-\boldsymbol{\psi}_u\|}, \ \ u\in\{{B},{E}\},
\end{equation}
where $\lambda$ is the carrier wavelength and $\eta$ denotes the effective antenna/waveguide gain. The BS transmits a confidential symbol $s$ with unit power, i.e., $\mathbb{E}\{|s|^2\}=1$, using a fixed transmit power $P_s$. The received signal at Bob is
\begin{equation}
    r_B=\sqrt{P_s}\,h_B\,s\,e^{-j\phi}+n_B,
\end{equation}
where $\phi = \frac{2\pi}{ \, \lambda_{g}}\|{\psi}_{p} - \psi_{0}\|$ denotes the phase shift incurred by the signal propagation inside the waveguide,
$\lambda_{g} =\lambda/n_{\mathrm{eff}}$ is the guided wavelength in the waveguide, and $n_B\sim \mathcal{CN}(0,\sigma^2)$. Therefore, Bob’s instantaneous SNR is
\begin{equation}
    \gamma_B = 
    \frac{\eta P_s}{\sigma^2\|\boldsymbol{\psi}_p-\boldsymbol{\psi}_{\text{Bob}}\|^2}
    =
    \frac{\bar{\gamma}}{E^2 + y_1^2 + h^2},
\end{equation}
where $\bar{\gamma}=\eta P_s/\sigma^2$ denotes the effective transmit SNR. 

Similarly, the received signal at Eve is
\begin{equation}
    r_E=\sqrt{P_s}\,h_E\,s\,e^{-j\phi}+n_E,
\end{equation}
where $n_E \sim \mathcal{CN}(0,\sigma^2)$. Due to pinching-position uncertainty, a horizontal mismatch is introduced between the radiating point and Eve, which affects the propagation distance \footnote{The locations of Bob and Eve directly affect the corresponding SNRs through the propagation distances and hence the resulting secrecy performance.}. Accordingly, the instantaneous SNR at Eve is given by
\begin{equation}
    \gamma_E=\frac{\eta P_s}{\sigma^2\|\boldsymbol{\psi}_p-\boldsymbol{\psi}_{\text{Eve}}\|^2}
    =
    \frac{\bar{\gamma}}{(x_1+E-x_2)^2+y_2^2+h^2}.
\end{equation}
The probability density function (PDF) of $\gamma_B$ and $\gamma_E$ can be obtained as \eqref{eq:fgammaB_final} and \eqref{eq:fgammaE_final}, respectively, shown at the top of next page. The details are provided in Appendix A.

\begin{figure*}[!t]
\begin{equation}
\small
f_{\gamma_B}(\gamma)=
\begin{cases}
\displaystyle
\frac{\pi}{2\Delta D}\frac{\bar{\gamma}}{\gamma^2}, 
& \frac{\bar{\gamma}}{\Delta^2+h^2} \le \gamma \le \frac{\bar{\gamma}}{h^2}, 
\\[1.2ex]
\displaystyle
\frac{\bar{\gamma}}{\Delta D\, \gamma^2}\,
\arcsin\!\left(\frac{\Delta}{\sqrt{\frac{\bar{\gamma}}{\gamma}-h^2}}\right), 
& \frac{\bar{\gamma}}{D^2/4+h^2} < \gamma < \frac{\bar{\gamma}}{\Delta^2+h^2}, 
\\[1.2ex]
\displaystyle
\frac{\bar{\gamma}}{\Delta D\, \gamma^2} 
\Bigg[
\arcsin\!\left(\frac{\Delta}{\sqrt{\frac{\bar{\gamma}}{\gamma}-h^2}}\right)
-
\arcsin\!\left(\sqrt{1-\frac{D^2/4}{\frac{\bar{\gamma}}{\gamma}-h^2}}\right)
\Bigg], 
& \frac{\bar{\gamma}}{\Delta^2+D^2/4+h^2} \le \gamma \le \frac{\bar{\gamma}}{D^2/4+h^2}, 
\\[1ex]
0, & \text{otherwise}.
\end{cases}
\label{eq:fgammaB_final}
\end{equation}
\vspace{0.2mm} 
\hrule  
\vspace{0.2mm} 
\end{figure*}

\begin{figure*}[!t]
\begin{equation}
\small
f_{\gamma_E}(\gamma)=
\begin{cases}
\displaystyle
\frac{\bar{\gamma}}{\gamma^2}
\frac{\pi}{2\Delta D^3}
\!\left(
2D\Delta-\Delta^2-\frac{\frac{\bar{\gamma}}{\gamma}-h^2}{2}
\right),
&
\!\!\frac{\bar{\gamma}}{h^2+\Delta^2}\le\gamma<\frac{\bar{\gamma}}{h^2},
\\[1.2ex]
\displaystyle
\frac{\bar{\gamma}}{\gamma^2}
\frac{1}{2\Delta D^3}
\!\left[
2\Delta D\pi
-\!\left(2\Delta^2+\frac{\bar{\gamma}}{\gamma}-h^2\right)
\arcsin\!\!\left(\frac{\Delta}{\sqrt{\frac{\bar{\gamma}}{\gamma}-h^2}}\right)
-3\Delta\sqrt{\frac{\bar{\gamma}}{\gamma}-h^2-\Delta^2}
\right],
&
\!\!\frac{\bar{\gamma}}{h^2+\frac{D^2}{4}}
\le\gamma<
\frac{\bar{\gamma}}{h^2+\Delta^2},
\\[1.2ex]
\displaystyle
\frac{\bar{\gamma}}{\gamma^2}
\,f_S^{(3)}\!\left(\frac{\bar{\gamma}}{\gamma}-h^2\right),
&
\!\!\frac{\bar{\gamma}}{h^2+(D+\Delta)^2+\frac{D^2}{4}}
\le\gamma<
\frac{\bar{\gamma}}{h^2+\frac{D^2}{4}},
\\[1ex]
0, & \text{otherwise}.
\end{cases}
\label{eq:fgammaE_final}
\end{equation}
\vspace{0.2mm} 
\hrule  
\vspace{0.2mm} 
\end{figure*}



The instantaneous secrecy rate of the system is defined as
\begin{equation}
    R_s=\left[\log_2(1+\gamma_B)-\log_2(1+\gamma_E)\right]^+,
\end{equation}
where $[x]^+ \triangleq \max\{x,0\}$. Since the BS lacks Eve’s instantaneous CSI, it transmits at a fixed secrecy rate $R_{\mathrm{th}}$. A secrecy outage occurs whenever the achievable secrecy rate falls below this target, i.e., $R_s < R_{\mathrm{th}}$.
Accordingly, the SOP is
\begin{equation}
    P_{\mathrm{out}}^{\mathrm{sec}}
    =
    \Pr\!\left(
        \log_2(1+\gamma_B)-\log_2(1+\gamma_E)
        \le R_{\mathrm{th}}
    \right),
\end{equation}
which can be equivalently expressed as
\begin{equation}
    P_{\mathrm{out}}^{\mathrm{sec}}
    =
    \Pr\!\left(
        \gamma_B < g(\gamma_E)
    \right),
\end{equation}
where $g(\gamma_E)
    \triangleq 2^{R_{\mathrm{th}}}\big(1+\gamma_E\big) - 1$.

\section{Secrecy Performance Analysis}
\label{sec:SOP_analysis}

In this section, we analyze the secrecy reliability of the considered PAS under pinching-position uncertainty. Unlike conventional secrecy models that assume statistically independent legitimate and eavesdropping links, the considered PAS architecture inherently introduces statistical dependence between $\gamma_B$ and $\gamma_E$ through the shared activation-position uncertainty and the underlying propagation geometry. This uncertainty-induced dependence invalidates the conventional independence assumption and requires a dependence-aware secrecy analysis. To characterize this dependence, the PAS-specific marginal SNR distributions derived in Section II are coupled through a Gaussian copula to construct the corresponding joint SNR distribution \cite{9900277,10453225}. The Gaussian copula provides a mathematically tractable representation of the resulting joint SNR distribution, while the dependence itself originates from the PAS geometry and the shared activation-position uncertainty. Since the adopted uncertainty model is zero-mean and symmetric, the induced dependence is expected to be predominantly symmetric without directional or asymmetric tail behavior, making the Gaussian copula a suitable and analytically tractable choice for the considered PAS uncertainty model.


The following theorem presents the SOP characterization.

\textbf{Theorem~1.}
{For the secure PAS with random pinching-position uncertainty, the SOP can be approximated as}
\begin{equation}
    P_{\mathrm{out}}^{\mathrm{sec}}
    \approx
    \frac{\pi}{N}
    \sum_{n=1}^{N}
        \sqrt{1-\xi_n^2}\,
        \Gamma(\xi_n),
    \label{eq_SOP_GCQ}
\end{equation}
where $N$ denotes the number of Gauss--Chebyshev nodes,
\begin{equation}
    \xi_n = \cos\!\left(\frac{2n-1}{2N}\pi\right), \quad n=1,\dots,N,
\end{equation}
and
\begin{align}
\small
    \Gamma(u)
    &\triangleq
    \Phi\!\left(
        \frac{
            \Phi^{-1}\!\big(F_{\gamma_B}(g(\gamma_E(u)))\big)
            - \rho\,\Phi^{-1}\!\big(F_{\gamma_E}(\gamma_E(u))\big)
        }{
            \sqrt{1-\rho^2}
        }
    \right)
 \nonumber\\
    &\quad\times
    f_{\gamma_E}(\gamma_E(u))
    \frac{\gamma_{E,\max} - \gamma_{E,\min}}{2},
\end{align}
with the affine mapping between $u\in[-1,1]$ and the support of $\gamma_E$ is 
$\gamma_{E}(u)=((\gamma_{E,\max}-\gamma_{E,\min})/2)u+(\gamma_{E,\max}+\gamma_{E,\min})/2$, 
with $\gamma_{E,\min}=\bar{\gamma}/\big(h^{2}+(D+\Delta)^{2}+D^{2}/4\big)$ and 
$\gamma_{E,\max}=\bar{\gamma}/h^{2}$. 
Moreover, $\Phi(\cdot)$ and $\Phi^{-1}(\cdot)$ denote the standard normal CDF and its inverse, respectively, and $\rho$ represents the Gaussian-copula correlation coefficient \cite{10453225} between $\gamma_B$ and $\gamma_E$.

\textit{Proof:} The detailed proof is provided in Appendix~B.

\textbf{Remark 1.} When the pinching-position uncertainty vanishes, i.e., $\Delta \to 0$, the deviation term vanishes and the pinching point is perfectly aligned with the legitimate receiver. As a result, the dependence between $\gamma_B$ and $\gamma_E$ disappears, and the SOP reduces to the conventional independence-based expression \cite{pinchplc2, 11215679}.

\section{Simulation Results}

This section presents numerical results to validate the derived analytical expressions and quantify the impact of uncertainty-induced dependence on the secrecy performance of PAS. Unless otherwise stated, the service area is a square of side $D = 20$~m, the waveguide height is $h = 5$~m, and the target secrecy rate is $R_{\mathrm{th}} = 0.5$~bit/s/Hz. The carrier frequency is $f_{c} = 28$~GHz with an effective refractive index of $n_{\mathrm{eff}} = 1.4$, and the receiver noise power is $\sigma^{2} = -90$~dBm. The Gaussian--Chebyshev quadrature used in Theorem~1 employs $N = 200$ nodes to ensure numerical stability, and MC simulations are generated from $10^4$ random realizations. For comparison, a conventional fixed-antenna configuration, where the BS radiates from a static location, is considered as a benchmark.\footnote{The fixed antenna is located at $(0,0,h)$ and operates under the same transmit power, propagation assumptions, carrier frequency, and effective SNR normalization as the proposed PAS scheme.}

\begin{figure*}[t] 
    \centering
    \subfloat[]{%
        \includegraphics[width=0.33\linewidth, height=0.17\linewidth]{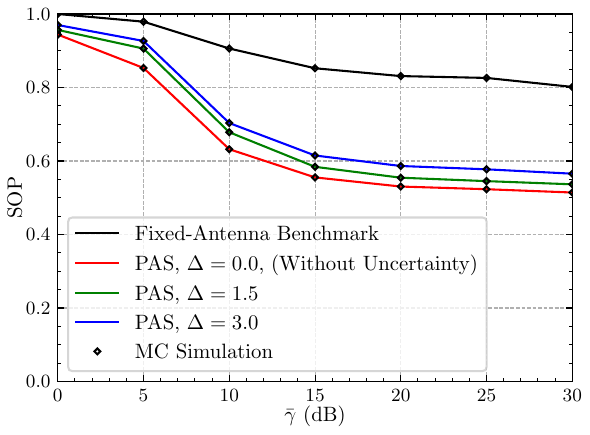}%
        \label{fig:fig1a}%
    }
    \hfill
    \subfloat[]{%
        \includegraphics[width=0.33\linewidth, height=0.17\linewidth]{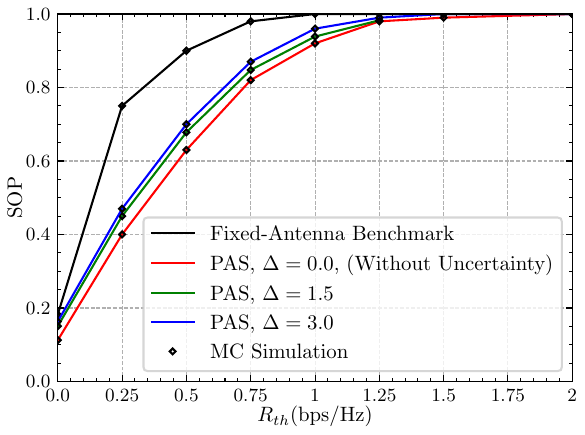}%
        \label{fig:fig1b}%
    }
    \hfill
    \subfloat[]{%
        \includegraphics[width=0.33\linewidth, height=0.17\linewidth]{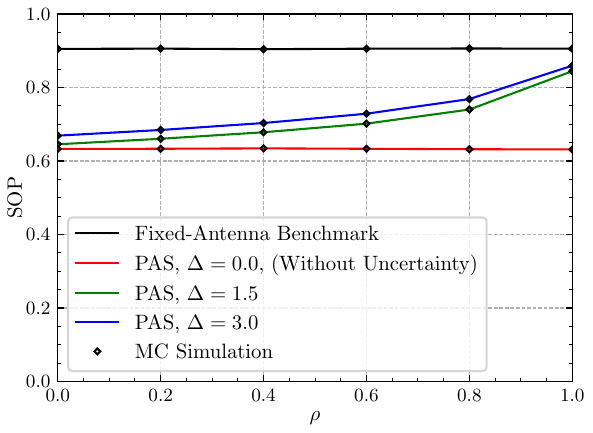}%
        \label{fig:fig1c}%
    }
    \caption{SOP performance of the considered secure PAS: 
(a) versus average transmit SNR, 
(b) versus target secrecy rate $R_{\mathrm{th}}$, and 
(c) versus correlation coefficient $\rho$, 
under different pinching-position uncertainty levels.}

    \label{fig:all_cases}
\end{figure*}

Fig.~2a depicts the SOP versus effective transmit SNR. The results demonstrate that the PAS operating under ideal pinching conditions consistently achieves a lower SOP than the fixed-antenna benchmark across all SNRs, thereby highlighting the inherent secrecy advantage enabled by spatial reconfigurability. Meanwhile, as the level of pinching-position uncertainty increases, SOP gradually deteriorates. This behavior stems from the loss of spatial control over the activated radiating point, which reduces the PAS ability to favor the legitimate receiver while simultaneously strengthening the uncertainty-induced dependence between the legitimate and eavesdropping links.

Fig.~2b shows the SOP versus the secrecy-rate threshold $R_{\mathrm{th}}$ at a fixed average SNR. As expected, increasing $R_{\mathrm{th}}$ imposes stricter secrecy requirements, leading to a gradual rise in the SOP across all considered schemes. The fixed-antenna system exhibits the highest outage levels, reflecting its limited ability to enhance legitimate reception or mitigating eavesdropping. In contrast, PAS under ideal pinching significantly reduces SOP by dynamically steering the radiating point toward the legitimate user. When pinching-position uncertainty is introduced, the SOP increases due to imperfect positioning of the radiating point. Consequently, the spatial advantage offered by PAS is partially reduced. Nevertheless, PAS consistently outperforms the fixed-antenna benchmark, indicating that its secrecy advantage remains robust even under practical activation-position uncertainty.


Fig.~2c depicts the SOP versus the correlation coefficient $\rho$ between the legitimate and eavesdropping channels under different uncertainty levels. As $\rho$ increases, SOP generally degrades, since stronger channel dependence enables the eavesdropper to benefit from favorable propagation conditions experienced by the legitimate receiver. As a result, the spatial discrimination capability of PAS is reduced, making it more difficult to favor Bob over Eve, particularly when the two receivers experience similar geometry-dependent propagation conditions. These results highlight that activation-position uncertainty affects secrecy performance not only through reduced spatial focusing, but also through the uncertainty-induced statistical dependence between the legitimate and eavesdropping links, underscoring the importance of explicitly accounting for uncertainty-induced dependence when evaluating the secrecy performance of practical PAS.


\section{Conclusion}

This paper investigated the secrecy performance of PAS under practical pinching-position uncertainty. By explicitly accounting for the resulting statistical dependence between the legitimate and eavesdropping links, closed-form expression for the SOP was derived using a dependence-aware analytical framework in which a Gaussian copula is employed to represent the induced joint statistics. Numerical results demonstrated that activation-position uncertainty changes the statistical structure of the secrecy problem by inducing dependence between the legitimate and eavesdropping links, highlighting the importance of dependence-aware secrecy analysis for practical PAS. Nevertheless, PAS consistently preserves a secrecy advantage over conventional fixed-antenna systems despite activation-position uncertainty. The proposed framework provides a foundation for future investigations incorporating more realistic propagation environments, advanced uncertainty models, and additional practical implementation impairments.


\appendices

\section{}
\label{appendixA}

\textit{PDF of $\gamma_B$:}
Define the nonnegative random variables $U=E^2$, $V=y_1^2$, and $W=U+V$. Since
$E\sim\mathrm{Unif}[-\Delta,\Delta]$ and $y_1\sim\mathrm{Unif}[-D/2,D/2]$, their PDFs are given by
\begin{equation}
    f_U(u)=\frac{1}{2\Delta\sqrt{u}}, \quad 0<u<\Delta^2,
\end{equation}
and
\begin{equation}
    f_V(v)=\frac{1}{D\sqrt{v}}, \quad 0<v<\frac{D^2}{4},
\end{equation}
respectively. Owing to the independence of $U$ and $V$, the PDF of $W$ is obtained via convolution as \cite{10976621}
\begin{equation}
    f_W(w)=
    \int_{\max(0,w-D^2/4)}^{\min(w,\Delta^2)}
    f_U(x)f_V(w-x)\,\mathrm{d}x,
\end{equation}
for $0<w<\Delta^2+D^2/4$. By using
$\int \left( 1/\sqrt{x(w-x)} \right)\mathrm{d}x
=2\arcsin\!\left(\sqrt{x/{w}}\right)$
and assuming $\Delta\le D/2$, a closed-form three-branch expression over
$[0,\Delta^2]$, $(\Delta^2,D^2/4]$, and $(D^2/4,\Delta^2+D^2/4]$ for $f_W(w)$ is obtained.

The instantaneous SNR at Bob is related to $W$ as
\begin{equation}
    \gamma_B=\frac{\bar{\gamma}}{W+h^2}
     \quad \Longleftrightarrow \quad
    W=\frac{\bar{\gamma}}{\gamma_B}-h^2,
\end{equation}
which is a strictly decreasing mapping. 
Applying the change-of-variable formula yields
\begin{equation}
    f_{\gamma_B}(\gamma)
    =
    \frac{\bar{\gamma}}{\gamma^2}
    f_W\!\left(\frac{\bar{\gamma}}{\gamma}-h^2\right),
    \ \ 
    \gamma_{B,\min}<\gamma<\gamma_{B,\max},
\end{equation}
where $\gamma_{B,\min}=\frac{\bar{\gamma}}{h^2+\Delta^2+D^2/4}$ and $\gamma_{B,\max}=\frac{\bar{\gamma}}{h^2}$.
Mapping the intervals of $w$ to those of $\gamma$ directly leads to the piecewise PDF in \eqref{eq:fgammaB_final}.

\textit{PDF of $\gamma_E$}: Let $X=x_1-x_2$ denote the horizontal separation between Bob and Eve. Since $x_1,x_2\sim\mathrm{Unif}[-D/2,D/2]$ are independent, $X$ has the triangular PDF
\begin{equation}
    f_X(x)
    =
    \begin{cases}
        \dfrac{D - |x|}{D^2}, & -D \le x \le D, \\[0.5ex]
        0, & \text{otherwise}.
    \end{cases}
\end{equation}
Defining $Z = X + E$, its PDF is obtained by convolution and exhibits a three-branch form
\begin{equation}
\small
f_Z(z) =
\begin{cases}
\dfrac{2D\Delta - \Delta^2 - z^2}{2\Delta D^2}, & |z| \le \Delta, \\[1ex]
\dfrac{D - |z|}{D^2}, & \Delta < |z| \le D - \Delta, \\[1ex]
\dfrac{(D + \Delta - |z|)^2}{4\Delta D^2}, & D - \Delta < |z| \le D + \Delta, \\[0.5ex]
0, & \text{otherwise}.
\end{cases}
\end{equation}
Define $U_Z=Z^2$ and $V_2=y_2^2$, where $y_2\sim\mathrm{Unif}[-D/2,D/2]$. Owing to symmetry, the PDF of $U_Z$ follows directly from that of $Z$, while $V_2$ has PDF
\begin{equation}
    f_{V_2}(v)=\frac{1}{D\sqrt{v}}, \quad 0<v<\frac{D^2}{4}.
\end{equation}
The squared-distance term in Eve’s SNR is $S=U_Z+V_2$, with support $0 \le S \le (D+\Delta)^2 + \frac{D^2}{4}$.

Since $U_Z$ and $V_2$ are independent, the PDF of $S$ is obtained by convolution. Partitioning the integral according to the regions of $f_{U_Z}(\cdot)$ yields a three-branch closed-form expression for $f_S(s)$, valid on the intervals $0 < s \le \Delta^2$, $\Delta^2 < s \le D^2/4$, and $D^2/4 < s \le (D+\Delta)^2 + \frac{D^2}{4}$. For the third interval, an exact closed-form expression is not provided due to space constraints, but it can be directly obtained via convolution.
Finally, the mapping between $S$ and $\gamma_E$ is
\begin{equation}
    \gamma_E
    =
    \frac{\bar{\gamma}}{S + h^2}
    \quad \Longleftrightarrow \quad
    S
    =
    \frac{\bar{\gamma}}{\gamma_E} - h^2,
\end{equation}
which is strictly decreasing in $S$. 
Applying the change-of-variable formula gives
\begin{equation}
    f_{\gamma_E}(\gamma)
    =
    \frac{\bar{\gamma}}{\gamma^2}\,
    f_S\!\left(\frac{\bar{\gamma}}{\gamma} - h^2\right),
    \ \ 
    \gamma_{E,\min}<\gamma<\gamma_{E,\max},
\end{equation}
where $\gamma_{E,\min}
    =
    \frac{\bar{\gamma}}{h^2 + (D+\Delta)^2 + \frac{D^2}{4}}$ and $\gamma_{E,\max}
    =
    \frac{\bar{\gamma}}{h^2}$. Mapping the intervals yields the piecewise PDF in
\eqref{eq:fgammaE_final}. \hfill $\blacksquare$

\appendices
\section*{Appendix~B}
\label{appendix:SOP_proof}

Starting from (11) and applying Sklar’s theorem \cite{9900277, 10453225}, the joint distribution of $(\gamma_B,\gamma_E)$ can be written as
\begin{equation}
    f_{\gamma_B,\gamma_E}(x,y)
    = c\!\left(F_{\gamma_B}(x),F_{\gamma_E}(y);\rho\right)
      f_{\gamma_B}(x) f_{\gamma_E}(y),
    \label{eq:app_jointPDF}
\end{equation}
where $c(\cdot,\cdot;\rho)$ denotes the Gaussian copula density, $\rho$ is its dependence parameter, and
$F_{\gamma_B}(\cdot)$, $F_{\gamma_E}(\cdot)$ are the PAS–induced marginal CDFs. Using (11) and \eqref{eq:app_jointPDF},
\begin{align}
    P_{\mathrm{out}}^{\mathrm{sec}}
    &=
    \iint_{x<g(y)}
        c\left(F_{\gamma_B}(x),F_{\gamma_E}(y);\rho\right)
        f_{\gamma_B}(x) f_{\gamma_E}(y)
    \,\mathrm{d}x\,\mathrm{d}y .
    \label{eq:app_SOP_double}
\end{align}
Introducing the probability integral transforms $U = F_{\gamma_B}(\gamma_B)$ and $V = F_{\gamma_E}(\gamma_E)$,
the variables $(U,V)$ are uniform on $(0,1)$ and retain the same Gaussian copula.
The conditional CDF of $U$ given $V=v$ is
$F_{U|V}(u|v)
    =
    \Phi\!\left(
        \frac{\Phi^{-1}(u) - \rho\,\Phi^{-1}(v)}
             {\sqrt{1-\rho^2}}
    \right)$, obtained from the conditional Gaussian law of $(Z_1|Z_2)$ in the normal representation of the copula.
Mapping back via $u=F_{\gamma_B}(x)$ and $v=F_{\gamma_E}(y)$ yields
\begin{equation}
    F_{\gamma_B|\gamma_E}(x|y)
    =
    \Phi\!\left(
        \frac{
            \Phi^{-1}\!\big(F_{\gamma_B}(x)\big)
            - \rho\,\Phi^{-1}\!\big(F_{\gamma_E}(y)\big)
        }{
            \sqrt{1-\rho^2}
        }
    \right),
    \label{eq:app_condCDF_original}
\end{equation}
which replaces the inner integral in \eqref{eq:app_SOP_double}.  
Hence,
\begin{equation}
    P_{\mathrm{out}}^{\mathrm{sec}}
    =
    \int_{\gamma_{E,\min}}^{\gamma_{E,\max}}
        F_{\gamma_B|\gamma_E}\big(g(y)\mid y\big)\,
        f_{\gamma_E}(y)\,
    \mathrm{d}y.
\end{equation}

To obtain the approximation in \eqref{eq_SOP_GCQ}, define the affine transformation
\begin{equation}
    y(u)
    =
    \frac{\gamma_{E,\max}-\gamma_{E,\min}}{2}\,u
    +
    \frac{\gamma_{E,\max}+\gamma_{E,\min}}{2},
    \ -1\le u\le 1.
\end{equation}
The SOP integral becomes
    $P_{\mathrm{out}}^{\mathrm{sec}}
    =
    \int_{-1}^{1} G(u)\,\mathrm{d}u$,
where $G(u)$ is the smooth transformation of the original integrand.  
By rewriting $\int_{-1}^{1} G(u)\,\mathrm{d}u
    =
    \int_{-1}^{1} \frac{\sqrt{1-u^2}\,G(u)}{\sqrt{1-u^2}}\,\mathrm{d}u$, and applying the $N$–point Gauss–Chebyshev quadrature of the first kind, the integral is approximated as (12).
This completes the proof.
\hfill $\blacksquare$

\bibliographystyle{IEEEtran}
\bibliography{refpinching}

@ARTICLE{8869705,
  author={W. Saad and others},
  journal={IEEE Netw.}, 
  title={A Vision of {6G} Wireless Systems: Applications, Trends, Technologies, and Open Research Problems}, 
month  = {Jun.},
  year={2020},
  volume={34},
  number={3},
  pages={134-142}}

@ARTICLE{10506508,
  author={A. Ashraf and others},
  journal={IEEE Access}, 
  title={Advancements and Challenges in Scalable Modular Antenna Arrays for {5G} Massive {MIMO} Networks}, 
month  = {Apr.},
  year={2024},
  volume={12},
  number={},
  pages={57895-57916}}

@ARTICLE{9900277,
  author={I. Trigui and others},
  journal={IEEE Trans. Veh. Technol.}, 
  title={Reconfigurable Intelligent Surfaces-Aided Relaying in Correlated {R}ice Fading: Performance Analysis and Design Optimization}, 
month  = {Apr.},
  year={2024},
  volume={73},
  number={4},
  pages={5150-5161}}

@ARTICLE{11106811,
  author={S. Pakravan and others},
  journal={IEEE Trans. Veh. Technol.}, 
  title={Fluid Antenna-Assisted Uplink {NOMA} Networks under Imperfect {SIC}}, 
year={2026},
  month  = {Jan.},
  volume={75},
  number={1},
  pages={1689-1694}
}

@ARTICLE{10945421,
  author={Z. Ding and others},
  journal={IEEE Trans. Commun.}, 
  title={Flexible-Antenna Systems: {A} Pinching-Antenna Perspective}, 
month     = {Oct.},
  year={2025},
  volume={73},
  number={10},
  pages={9236-9253}}

@ARTICLE{pinchplc1,
  author={G. Zhu and others},
  journal={IEEE Trans. Commun.}, 
  title={Pinching-Antenna Systems ({PASS})-Enabled Secure Wireless Communications},
month     = {Oct.},
  year={2025},
  volume={74},
  number={},
  pages={490-505}}

@article{pinchplc2,
  author    = {N. Li and others},
  title     = {On the Secrecy Performance of Pinching-Antenna Systems},
  journal   = {arXiv preprint arXiv:2509.16854},
  month     = {Sep.},
  year      = {2025}
}

@ARTICLE{11215679,
  author={Wang, Kaidi and others},
  journal={IEEE Wireless Commun. Lett.}, 
  title={Pinching-Antenna Systems for Physical Layer Security}, 
  year={2025},
  month     = {Oct.},
  volume={15},
  number={},
  pages={260-264}}

@ARTICLE{10896748,
  author={Y. Xu and others},
  journal={IEEE Wireless Commun. Lett.}, 
  title={Rate Maximization for Downlink Pinching-Antenna Systems}, 
month     = {May.},
  year={2025},
  volume={14},
  number={5},
  pages={1431-1435}}

@ARTICLE{pinch1111,
  author={M. Zing and others},
  journal={IEEE Wireless Commun. Lett.}, 
  title={Energy-Efficient Resource Allocation for {NOMA}-Assisted Uplink Pinching-Antenna Systems}, 
 month     = {Aug.},
  year={2025},
  volume={14},
  number={11},
  pages={3695-3699}}

@ARTICLE{pinchISAC,
  author={Y. Qin and others},
  journal={IEEE Wireless Commun. Lett.}, 
  title={Joint Antenna Position and Transmit Power Optimization for Pinching Antenna-Assisted {ISAC} Systems}, 
 month     = {Aug.},
  year={2025},
  volume={14},
  number={11},
  pages={3535-3539}}

@ARTICLE{10912473,
  author={K. Wang and others},
  journal={IEEE Wireless Commun. Lett.}, 
  title={Antenna Activation for {NOMA} Assisted Pinching-Antenna Systems}, 
month     = {May.},
  year={2025},
  volume={14},
  number={5},
  pages={1526-1530}}

@ARTICLE{11029492,
  author={Z. Zhou and others},
  journal={IEEE Wireless Commun. Lett.}, 
  title={Sum-Rate Maximization for {NOMA}-Assisted Pinching-Antenna Systems}, 
month     = {Sep.},
  year={2025},
  volume={14},
  number={9},
  pages={2728-2732}}

@ARTICLE{10976621,
  author={D. Tyrovolas and others},
  journal={IEEE Trans. Cogn. Commun. Netw.}, 
  title={Performance Analysis of Pinching-Antenna Systems}, 
month     = {Apr.},
  year={2025},
  volume={12},
  number={},
  pages={590-601}}

@ARTICLE{10453225,
  author={D. Shahbaztabar and others},
  journal={IEEE Open J. Commun. Soc.}, 
  title={Performance Analysis of {RIS}-Assisted Communication With Direct Link: A New Copula Application}, 
month     = {Feb.},
  year={2024},
  volume={5},
  number={},
  pages={1740-1752}}

@ARTICLE{11016750,
  author={Xie, Ximing and others},
  journal={IEEE Commun. Lett.}, 
  title={A Low-Complexity Placement Design of Pinching-Antenna Systems}, 
  year={2025},
  month     = {Aug.},
  volume={29},
  number={8},
  pages={1784-1788}}

@ARTICLE{10713386,
  author={Chen, Huashu and others},
  journal={IEEE Commun. Lett.}, 
  title={Energy-Efficient Wireless Communication With Irregular Reconfigurable Intelligent Surfaces}, 
  year={2024},
  month     = {Dec.},
  volume={28},
  number={12},
  pages={2754-2758}}

@article{suzuki2022pinching,
  author    = {Hiroshi Okazaki Yasunori Suzuki and other},
  title     = {Pinching Antenna: Using a Dielectric Waveguide as an Antenna},
  journal   = {NTT DOCOMO Tech. J.},
  volume     = {23},
  number     = {3},
  pages      = {5--12},
  year       = {2022},
  month      = {Jan}
}

@inproceedings{6363891,
	title={38 {GHz} and 60 {GHz} angle-dependent propagation for cellular \& peer-to-peer wireless communications},
	author={Rappaport, Theodore S. and others},
	booktitle={Proc. IEEE ICC},
	month={},
	pages={},
	year={},
note={{O}ttawa, {Canada}, pp. 4568--4573, Jun. 2012.}
}

@ARTICLE{10906511,
  author={Zhu, Lipeng and others},
  journal={IEEE Commun. Surv. Tutor.}, 
  title={A Tutorial on Movable Antennas for Wireless Networks}, 
  year={2026},
  month={Feb.},
  volume={28},
  number={},
  pages={3002-3054}}

@ARTICLE{10278220,
  author={Zhu, Lipeng and others},
  journal={IEEE Commun. Lett.}, 
  title={Movable-Antenna Array Enhanced Beamforming: Achieving Full Array Gain With Null Steering}, 
  year={2023},
  month={Oct.},
  volume={27},
  number={12},
  pages={3340-3344}}

@ARTICLE{11481959,
  author={Pakravan, Saeid and others},
  journal={IEEE Trans. Veh. Technol.}, 
  title={{AI}-Empowered Resource Allocation for Wirelessly Powered Pinching-Antenna Systems}, 
  year={2026},
  month={Apr.},
  volume={},
  number={},
  pages={1-6}}

\vfill
	
\end{document}